\documentclass[prd,amssymb,twocolumn,10pt]{revtex4-1}
\usepackage{amsmath,amsfonts,amssymb}
\usepackage{color,verbatim,graphicx,float}

\begin{document}

\newcommand{\dR}{\mathbb R}
\newcommand{\dC}{\mathbb C}
\newcommand{\dZ}{\mathbb Z}
\newcommand{\id}{\mathbb I}
\newcommand{\dT}{\mathbb T}

\title{Gaussian state for the bouncing quantum cosmology}

\author{Jakub Mielczarek and W{\l}odzimierz Piechocki
\\ Department of Fundamental Research,\\ National Centre for Nuclear Research,\\ Ho{\.z}a 69,
PL-00-681 Warsaw, Poland}

\date{\today}

\begin{abstract}
We present results concerning propagation of the Gaussian state
across the cosmological quantum bounce.  The reduced phase space
quantization of loop quantum cosmology is applied to the
Friedman-Robertson-Walker universe with a free massless
scalar field. Evolution of quantum moments of the canonical
variables is investigated. The covariance turns out to be a
monotonic function so it may be used as an evolution parameter
having quantum origin. We show that for the Gaussian state the
Universe is least quantum at the bounce. We propose explanation of
this counter-intuitive feature using the entropy of squeezing. The
obtained  time dependence of entropy is in agreement with
qualitative predictions based on von Neumann entropy for mixed
states. We show that, for the considered Gaussian state,
semiclassicality is preserved across the bounce, so there is no
cosmic forgetfulness.

\end{abstract}


\maketitle

\section{Introduction}

The available data of observational cosmology indicate that the
Universe emerged  from a state with extremely high energy
densities of matter fields. Theoretical cosmology,  in particular
the Belinskii-Khalatnikov-Lifshitz (BKL) scenario (generic
solution of general relativity  devoid of any assumption on the
symmetry of spacetime) predicts an existence of the initial
cosmological singularity with diverging gravitational and matter
fields invariants \cite{BKL1,BKL2,BKL3}. Both the observation and
the theory indicate that  the cosmological singularity, identified
at the classical level, may have much to do with an extreme
initial state in the evolution of the Universe. An existence of
the general solution to the Einstein equations with the
cosmological singularity means that this classical theory is
incomplete. It is expected that finding the singularity free
quantum BKL theory would help in the construction of a
theory unifying gravitation and quantum physics, and could be used
to describe the very early Universe.

The challenge of quantization of the BKL scenario should be
preceded by complete understanding of quantum aspects of some
special cases of this model. For instance, the
Friedman-Robertson-Walker (FRW) model can be used for such a
purpose \cite{OIB}. In particular, an {\it evolution} of the
quantum FRW model has not been fully understood yet.

Presently, quantum cosmology effects are commonly addressed within
loop quantum cosmology (LQC) methods. This approach has two
alternative forms: standard LQC (see, e.g.,
\cite{Ashtekar:2003hd,Bojowald:2006da,Ashtekar:2006wn}) and
nonstandard LQC (see, e.g.,
\cite{Dzierzak:2009ip,Malkiewicz:2009qv,Mielczarek:2011mx}). The
LQC methods rely on modifying general relativity by approximating
the curvature of connection by holonomies around small loops. Such
modification leads to replacing classical singularities by quantum
bounces \cite{Ashtekar:2003hd,Bojowald:2006da,Ashtekar:2006wn,
Dzierzak:2009ip,Malkiewicz:2009qv,Mielczarek:2011mx}.

In this paper, we apply the nonstandard LQC to the flat FRW model
coupled to a free massless scalar field.  The presented results
are a continuation of investigations initiated in our recent paper
\cite{Mielczarek:2011mx}. Here, we focus on analysis of the
quantum dynamic for the state being a Gaussian packet
constructed by using eigenstates of a physical
Hamiltonian. There are at least two reasons to consider this
particular state. First, strictly analytical results can be
obtained  allowing  detailed analysis of the model. It is
advantageous to have analytical toy models for testing  more
sophisticated models.  Second, the Gaussian packets are shown to
emerge in the process of decoherence \cite{Zurek:1991vd,Zurek:2003zz}
for various physical systems (see, e.g., \cite{Zurek:1992mv, Kiefer:2006je}).
Therefore, they serve as a good description of semiclassicality. The
minisuperspace model of the Universe can be treated as an outcome
of decoherence of the relevant degrees of freedom (as the scale factor)
with respect to the irrelevant degrees of freedom (as perturbations) belonging
to superspace. Thus, one can expect the minisuperspace model of
the Universe in the semiclassical regime to be described by the
Gaussian-type packet. Moreover, even if the obtained state is not
exactly the Gaussian function, the Gaussian one serves as a
reasonable approximation of any other single-peaked distribution.

The organization  of the paper is as follows. In Sec. II, we
introduce the quantum Hamiltonian for the considered model and  study
the corresponding eigenproblem, which was solved in Ref.
\cite{Mielczarek:2011mx}. The obtained eigenfunctions are
subsequently used to construct the Gaussian packet state. In Sec.
III, we study the evolution of the quantum moments of the canonical
variables for the state under consideration. We focus on
investigating dynamics of mean values, dispersions, and covariance.
Our analysis shows that for the Gaussian packet, quantum
uncertainty is minimal at the  bounce.  In Sec. IV, notion of
quantum entropy of squeezing is introduced. Evolution of this
entropy for the considered state is investigated. We show that the
minimum uncertainty at the bounce is related with the state of
minimum entropy.  In Sec. V, relative fluctuations are studied in
the context of the so-called cosmic forgetfulness. In  Sec. VI, we
draw conclusions and indicate further directions of investigation.

\section{Hamiltonian and evolution}

The {\it physical} classical Hamiltonian of the system is found to
be \cite{Mielczarek:2011mx}
\begin{equation}\label{cH}
H_{\lambda} = \frac{2}{\lambda\sqrt{G}} P \sin(\lambda Q),
\end{equation}
where $G$ is Newton's constant. The parameter $\lambda$ of the
theory may be related to the minimum area of the loop used to
determine the holonomies (its precise value in the nonstandard LQC
has to be fixed observationally). The variables $Q$ and $P$ may be
interpreted in terms of the Hubble constant and the volume of
space, respectively. They satisfy the algebra $\{Q,P\}=1$, and
have the domains $\lambda Q\in [0,\pi]$ and $P> 0$.

The {\it quantum} Hamiltonian corresponding to (\ref{cH}) reads
\cite{Mielczarek:2011mx}
\begin{equation}\label{eH}
\hat{H}_{\lambda}\psi = -\frac{i}{\lambda \sqrt{G}}\left(
2\sin(\lambda Q) \frac{d}{dQ}+\lambda \cos(\lambda Q) \right)\psi,
\end{equation}
where $\psi \in L^2([0,\pi/\lambda],dQ)=: \mathcal{H}$. The
eigenvalue problem  $\hat{H}_{\lambda}\Psi_E = E \Psi_E$ has the
solution
\begin{equation}\label{eigen}
\Psi_E(x)  =  \sqrt{ \frac{\lambda \sqrt{G}}{4\pi} \cosh\ \left(
\frac{2}{\sqrt{G}} x\right) }\; e^{iEx},~~~~E\in \dR ,
\end{equation}
where $x:= \frac{\sqrt{G}}{2} \ln \left| \tan\left( \frac{\lambda
Q}{2} \right) \right|$. One may easily verify that $\langle \Psi_E
| \Psi_{E'} \rangle = \delta(E'-E)$.

We specify the domain of the unbounded operator
$\hat{H}_{\lambda}$ as follows:
\begin{equation}
\label{dom}
D(\hat{H}_{\lambda}):= {\rm span} \{ \varphi_k,~~k \in
\mathbb{Z} \},
\end{equation}
where
\begin{equation}
\varphi_k(Q):=
\int_{-\infty}^\infty  c_k (E)
    \Psi_{E}(Q)\; dE,~~~c_k \in C^\infty_0 (\dR).
\end{equation}
The domain $D(\hat{H}_{\lambda})$ is a dense subspace of
$\mathcal{H}$, and an action $\hat{H}_{\lambda}$ does not lead
outside of $D(\hat{H}_{\lambda})$. It has been shown in
\cite{Mielczarek:2011mx} that $\hat{H}_{\lambda}$ is an
essentially {\it self-adjoint} operator on
$D(\hat{H}_{\lambda})$.

Making use of the Stone theorem \cite{RaS}, we define the {\it
unitary} operator of an  evolution as follows:
\begin{equation}\label{evol}
\hat{U}(s) := \exp\left\{- \frac{i}{\hslash} s \hat{H}_{\lambda}  \right\},
\end{equation}
where $s \in \dR$ is  a ``time" parameter. The state at any moment
of time, \cite{conv}, can be found as follows: $| \Psi(s) \rangle
= \hat{U}(s)| \Psi(0) \rangle = \exp\{- i s
 \hat{H}_{\lambda} \}| \Psi(0) \rangle$.

Let us consider a superposition of the Hamiltonian eigenstates $|
\Psi(0) \rangle = \int_{-\infty}^{+\infty} dE c(E) | \Psi_{E}
\rangle$ at $s=0$. Then, evolution of this state is given by
\begin{equation}\label{sup}
|\Psi(s)\rangle
=\int_{-\infty}^{+\infty}  dE c(E) e^{- i s
E}|\Psi_{E}\rangle .
\end{equation}
In what follows, we consider the Gaussian packet with a simple
profile,
\begin{equation}\label{prof}
c(E) := \sqrt[4]{ 2\alpha/\pi } \exp\left\{-\alpha
(E-E_0)^2\right\},
\end{equation}
that is centered at $E_0$ with the dispersion parametrized by
$\alpha$. We find that the normalized packet defined by
(\ref{sup}) and (\ref{prof}) reads
\begin{equation}
\label{state}
\Psi(x,s) =  \sqrt{\frac{\lambda \cosh\ \left( \frac{2}{\sqrt{G}} x\right)}
{\sqrt{8\pi \tilde{\alpha}}}}
e^{-\frac{(x-s)^2}{4\alpha}}e^{i E_0(x-s)},
\end{equation}
where $\tilde{\alpha} := \alpha/G$ (for further purposes, we also
define $\tilde{E}_0 := E_0\sqrt{G}$).

\section{Quantumness}

To get some insight into the {\it nature} of the quantum bounce,
one studies possible correlation between quantum fluctuations
before and after the bounce. In particular, one tries to find the
answer to the question:  Is the  semiclassicality of the Universe
preserved across the bounce? If the answer is negative, we
would not be able to learn what had happened before the big
bounce. It is called the cosmic {\it forgetfulness} or amnesia
\cite{BojowaldNature,Corichi:2007am}. We will come to this problem
in Sec. V.

Another issue, partially related with forgetfulness, is the {\it
quantumness} of the Universe. This property can be studied by
analyzing quantum moments of the canonical variables. Here, we
will focus on mean values $\langle \hat{Q} \rangle$ and $\langle
\hat{P} \rangle$, dispersions $\Delta \hat{Q}:= \sqrt{\langle
\hat{Q}^2 \rangle-\langle \hat{Q} \rangle^2 }$ and $\Delta
\hat{P}:=\sqrt{\langle \hat{P}^2 \rangle-\langle \hat{P} \rangle^2
}$, and covariance
\begin{eqnarray}
C_{QP} &:=& \langle (\hat{Q}-\langle \hat{Q}\rangle)(\hat{P}-\langle\hat{P}\rangle )\rangle \nonumber \\
&=& \frac{1}{2}\langle \hat{Q} \hat{P}+\hat{P}\hat{Q} \rangle - \langle \hat{Q} \rangle\langle \hat{P} \rangle.
\end{eqnarray}
The lowest moments listed above give us the basic characteristics
of a quantum state.

\begin{figure}[ht!]
\centering
\includegraphics[width=7cm,angle=0]{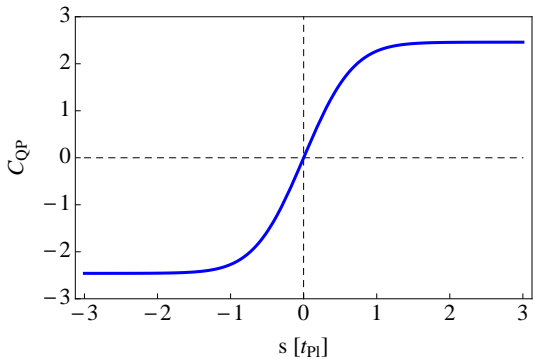}
\caption{Covariance $C_{QP}$ for $\tilde{\alpha}=0.1$ and $\tilde{E}_0=10$.}
\label{CovQP}
\end{figure}

An important property of the quantum state (\ref{state}) is that
its covariance is nonvanishing. In Fig. \ref{CovQP}, we show
a representative example of $C_{QP}$ as a function of time.
The covariance is a monotonic function of time, and it crosses
zero value just at the point of bounce ($s=0$). The monotonic
behavior of $C_{QP}$ suggest that covariance may serve as an
internal time parameter. In such a case, the time would have purely
quantum nature, different from the time parameter $s$ used in
(\ref{evol}). The idea that covariance may play a role of time was
already proposed in Ref. \cite{Bojowald:2008by} and later explored
in Ref. \cite{Bojowald:2009kb}. It was shown there that in case of
recollapsing  cosmology, the covariance is also a monotonic
function. Whether the perceptible flow of time has the purely
quantum origin or not remains an open issue. We will come to this
problem in the following section when considering quantum entropy
of squeezing.

\begin{figure}[ht!]
\centering
\includegraphics[width=7cm,angle=0]{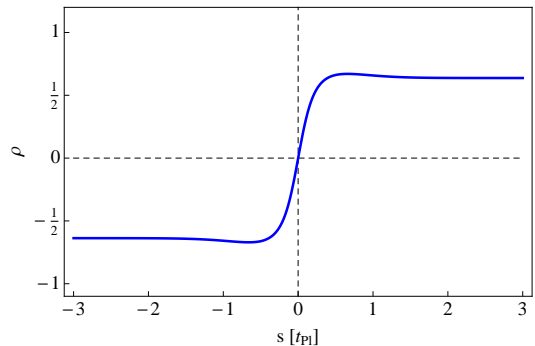}
\caption{Correlation coefficient $\rho$ for $\tilde{\alpha}=0.1$ and $\tilde{E}_0=10$.}
\label{CorCoef}
\end{figure}

The  covariance $C_{QP}$  together with dispersions $\Delta \hat{Q}$
and $\Delta \hat{P}$ can be utilized to define the dimensionless correlation coefficient
\begin{equation}
\rho :=  \frac{C_{QP}}{\Delta \hat{Q} \Delta \hat{P}}.
\end{equation}
We show time dependence of this coefficient in Fig. \ref{CorCoef}.
Similarly, like for the case of covariance, the correlation
coefficient $\rho$ saturates for $s\rightarrow \pm \infty$.
Therefore, one can infer that also a product of uncertainties
$\Delta \hat{Q}\Delta \hat{P}$ approaches asymptotically some
finite, nonzero values.
 Furthermore, it is worth mentioning that in contrast to  $C_{QP}$, the correlation coefficient
$\rho$ is not a monotonic function of $s$, so it cannot play the 
role of the intrinsic quantum parameter of time.

\begin{figure}[ht!]
\centering
\includegraphics[width=7cm,angle=0]{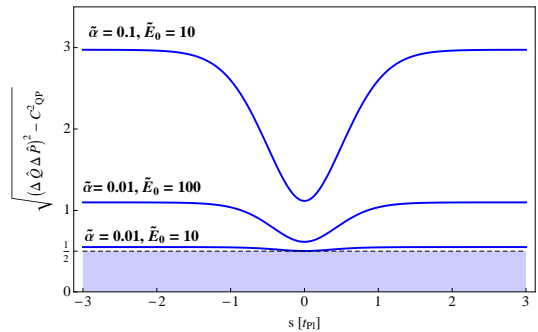}
\caption{Uncertainty $\sqrt{(\Delta\hat{Q})^2 (\Delta
\hat{P})^2-C_{QP}^2  }$ for $\tilde{\alpha}=0.1$ and
$\tilde{E}_0=10$.  The shadowed region is prohibited in
accordance with the uncertainty principle (\ref{RSrelation}).}
\label{Uncertainty}
\end{figure}

For a state with nonvanishing covariance, as the one studied here, the
so-called Robertson-Schr{\"o}dinger uncertainty relation holds
\cite{Schrodinger:1930ty},
\begin{equation}
(\Delta\hat{Q})^2 (\Delta \hat{P})^2-C_{QP}^2  \geq (\hslash/2)^2,
\label{RSrelation}
\end{equation}
which is a generalization of the Heisenberg uncertainty relation.
The uncertainty relation is saturated, while the square of the 
left-hand side of the inequality (\ref{RSrelation}) reaches the
$\hslash/2$ value. Such a state of minimal uncertainty, usually
corresponding to the vacuum state, is the least quantum state.
In order to study how such state is approached within our model,
we plot the square of the left-hand side of (\ref{RSrelation}) as a
function of time in Fig. \ref{Uncertainty}.
As we see, uncertainty reaches its minimum  at the transition
point between the contracting and expanding phases. Therefore,
bounce can be viewed as the least quantum part of the evolution.
The Universe can be best ``localized" just at the bounce. In the
next section, we will try to explain this counter-intuitive feature
using entropy.

We have studied so far an evolution of covariance and dispersions
for our model. Let us now combine this knowledge with evolution of
mean values of the canonical variables and investigate evolution
of the quantum system under consideration on the phase space.  For
this purpose, let us consider the covariance matrix,
\begin{equation}
{\bf \Sigma} :=  \left[\begin{array}{cc} ( \Delta \hat{Q})^2 & C_{QP} \\ C_{QP}& ( \Delta \hat{P})^2\end{array}  \right],
\end{equation}
which is known in statistics. Since the ${\bf \Sigma}$ matrix is
symmetric, it can be diagonalized ${\bf \Sigma} = {\bf O} {\bf \Lambda} {\bf O}^{T}$,
where ${\bf O}$ is the rotation matrix and the diagonalized matrix ${\bf \Lambda}
=\text{diag}(\lambda_{+},\lambda_{-})$. The eigenvalues of
matrix ${\bf \Lambda}$ can be expressed as follows:
\begin{eqnarray}
\lambda_{\pm} = \frac{1}{2}\left[ \text{tr}  {\bf \Sigma} \pm \sqrt{(\text{tr}{\bf \Sigma})^2 -4 \det {\bf \Sigma} } \right],
\end{eqnarray}
where $\text{tr}  {\bf \Sigma}= (\Delta \hat{Q})^2 +( \Delta
\hat{P})^2$ and $\det {\bf \Sigma}=( \Delta \hat{Q})^2(\Delta
\hat{P})^2(1-\rho^2)$. The square roots of $\lambda_{+}$ and
$\lambda_{-}$ have interpretation of major and minor axes of the
ellipsoid of covariance, respectively. An angle between the major
axis and the $P$ axis (counting counterclockwise) is
\begin{equation}
\theta = \frac{1}{2}\arctan \left[  \frac{2 C_{QP}}{(\lambda \Delta \hat{Q})^2 -(\Delta \hat{P}/\lambda)^2}\right].
\end{equation}
It is worth noting that the angle $\theta$, in contrast to the
covariance $C_{QP}$ and the coefficient $\rho$, is an explicit
function of the parameter $\lambda$.

\begin{figure}[ht!]
\centering
\includegraphics[width=6cm,angle=0]{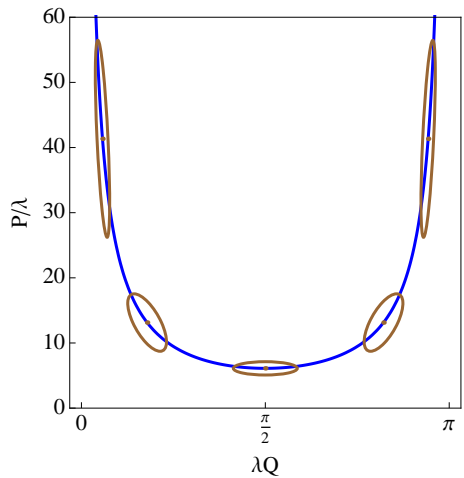}
\caption{ An open curve corresponds to the plot
($\langle\hat{Q}\rangle \lambda$,$\langle\hat{P}\rangle/\lambda$)
for $\tilde{\alpha}=0.1$ and $\tilde{E}_0=10$.  Ellipses describe
dispersion of state at five stages of evolution.} \label{Fig1}
\end{figure}

In Fig. \ref{Fig1} we show the parametric curve
($\langle\hat{Q}\rangle \lambda$, $\langle\hat{P}\rangle/\lambda$)
for $\tilde{\alpha}=0.1$ and $\tilde{E}_0 =10$ together
with five exemplary ellipses of dispersion.
The dispersion of $\hat{Q}$ is largest at the bounce and tends to
zero with the increase of $\langle \hat{P} \rangle$. In turn, the
dispersion $\Delta\hat{P}$ reaches its minimum value at the bounce
and grows with the increase of $\langle \hat{P} \rangle$.
Furthermore, as can be inferred from the behavior of the ellipses
of covariance, the state undergoes both amplitude and phase
squeezing.  This reflects the fact that both axes of the ellipse
of covariance ($\sqrt{\lambda_{\pm}}$) as well as the phase factor
$\theta$ are evolving in time.

\section{Entropy}

It is tempting to define the notion of entropy in quantum
cosmology as it may be used in  defining the cosmological arrow of
time.  The von Neumann definition of entropy for a quantum system
reads $S = -k_B \text{tr}[ \hat{\rho} \ln \hat{\rho}]$. Here,
$k_B$ is the Boltzmann constant and $\hat{\rho}$ is the density matrix
for a quantum state.  The von Neumann entropy is defined in such a
way that it vanishes  for the pure states $|\Psi \rangle$, for
which $\hat{\rho} = | \Psi \rangle \langle \Psi |$. One argues
that since an  evolution of the pure state is unitary, there is no
information loss so the entropy remains unchanged. On the other
hand, for mixed states, information about the quantum system can be
lost due to the entanglement with environment and the von Neumann
entropy is increasing. In particular, in quantum cosmology,
degrees of freedom can be decomposed for relevant and irrelevant
as already mentioned in the Introduction.   The irrelevant  degrees of
freedom may play a role of the environment that causes decoherence
of the sector parametrized by the relevant degrees of freedom as
the scale factor.  Making some qualitative analysis of the Wheeler-DeWitt
equation one may show that the resulting von Neumann entropy is
a monotonically increasing function of a scale factor for the FRW
model \cite{Zeh,Kiefer:2005vd}. This is in agreement with
expectations based on thermodynamical arguments for the expanding
Universe. Moreover, such dependence indicates that quantum entropy
should decrease  in the contracting Universe. This may suggest, by
applying the second law of thermodynamics, that the arrow of time is
directed out of the bounce and only expansion has physical
meaning \cite{Kiefer:2005vd}.

In the case discussed above, the entropy production was due to the
entanglement with the environment containing irrelevant degrees of
freedom. However, while considering the minisuperspace model of
quantum cosmology, as the one studied in this paper,  one usually
does not refer to the whole superspace and therefore no entropy
growth is expected, as for the pure states. This is however not
physically true and results from oversimplification of the system
under consideration.  However, we suppose that the information
about an entropy can be extracted from the state of the
minisuperspace model. This is because a form of this state is
determined by the process of decoherence, which takes into account
the environmental degrees of freedom. As already mentioned in the 
Introduction the resulting states are usually Gaussian packets, as
the one considered here.

Based on the above information, we derive the conclusion that the other 
definition of entropy should be used in such a case, which may give 
nonvanishing entropy also for some pure states. In fact, such an idea is 
not a new one. Especially, the relation between the degree of squeezing 
of a quantum state and entropy was discussed \cite{Gasperini:1993mq}.
It was shown that in the case of quantum cosmological
perturbations, the entropy of the squeezing pure semiclassical
states and the von Neumann entropy may lead to the same results
\cite{Gasperini:1995yd}. An idea of relating quantum evolution for
pure states with the arrow of time and entropy was also discussed by
Bojowald \emph{et al.} \cite{Bojowald:2008by,Bojowald:2009kb}.

We propose the following definition of entropy measuring the degree of
squeezing of a quantum state:
\begin{equation}
S := k_B \ln\left(\frac{ \Delta \hat{Q} \Delta \hat{P} \sqrt{1-\rho^2} }{\hslash/2}\right).
\label{DefEntropy}
\end{equation}
This definition  can be generalized to the cases with a higher
number of degrees of freedom.  Definition (\ref{DefEntropy}) may
be viewed as a quantum analogue of the Boltzmann entropy $S = k_B
\ln \Omega$, where $\Omega$ is the number of microstates in the
microcanonical ensemble. The minimal volume of phase space, which
can be occupied by the quantum system is obtained by saturating the 
Heisenberg  (or Robertson-Schr{\"o}dinger) relation so it is equal
to $\Gamma_0 = \hslash/2$. Therefore, $\Gamma/\Gamma_0$, where
$\Gamma:=  \Delta \hat{Q} \Delta \hat{P}\sqrt{1-\rho^2}
=\sqrt{\lambda_+\lambda_-}$ ($\sim$ area of ellipse of
covariance), is approximately the number of elementary cells
$\Gamma_0$  covered by the ellipsoid of covariance. This number of
cells is an analogue of the number of microstates $\Omega$ in
a given macrostate.

One can also show that equation (\ref{DefEntropy}) can be derived from the
Gibbs formula. Taking into account only first and second order quantum moments
to describe our system, the Wigner function can be approximated by the Gaussian
distribution
\begin{equation}
W(Q,P) = \frac{1}{2\pi\sqrt{\det {\bf \Sigma}}} \exp \left( -\frac{1}{2} {\bf x}^T {\bf \Sigma}^{-1} {\bf x} \right)
\label{Wigner}
\end{equation}
where ${\bf x} = (Q - \langle \hat{Q}  \rangle,  P - \langle \hat{P}  \rangle )$. The Wigner
function is strictly a positive function here, reflecting semiclassical nature of the state
under considerations. Moreover, the Wigner function fulfills the normalization condition
\begin{equation}
\int_{-\infty}^{+\infty} dQ   \int_{-\infty}^{+\infty} dP\  W(Q,P) =1.
\end{equation}
For the sake of simplicity,  we have extended here ranges of $Q$ and $P$
variables to $\mathbb{R}$. This approximation is however well satisfied
for the considered Gaussian state. Namely, by looking at Fig. \ref{Fig1},
we see that the ellipses of dispersions are placed well within the region
$[0,\pi]\times\mathbb{R}_+$. Therefore, contributions from the region
outside of $[0,\pi]\times\mathbb{R}_+$ to the Wigner function (\ref{Wigner})
is marginal.

Using the Wigner function (\ref{Wigner}) as a phase space density distribution
in definition of the Gibbs entropy we obtain
\begin{eqnarray}
S&=&-k_{B}  \int_{-\infty}^{+\infty} dQ   \int_{-\infty}^{+\infty} dP\ W(Q,P) \ln(W(Q,P))\nonumber \\
&=& k_{B}(1+\ln \pi)+k_{B} \ln\left(\frac{ \sqrt{\det {\bf \Sigma} }}{\hslash/2}\right)\nonumber \\
&=& k_{B}(1+\ln \pi)+k_B \ln\left(\frac{ \Delta \hat{Q} \Delta \hat{P} \sqrt{1-\rho^2} }{\hslash/2}\right),
\end{eqnarray}
where we restored the Planck constant $\hslash$.
So, the Gibbs entropy leads to formula (\ref{DefEntropy}) up to the constant factor $k_{B}(1+\ln \pi)$.
It is worth stressing that it was possible to use here the Wigner function, only because it was positive
for the state under consideration. However, in case of not strictly positive Wigner functions, one can
still construct the corresponding Husimi $Q$ function being positively defined. In this case, the above
Gibbs formula applied to the Husimi distribution leads to so-called Wehrl entropy \cite{Wehrl:1978zz}.

\begin{figure}[ht!]
\centering
\includegraphics[width=7cm,angle=0]{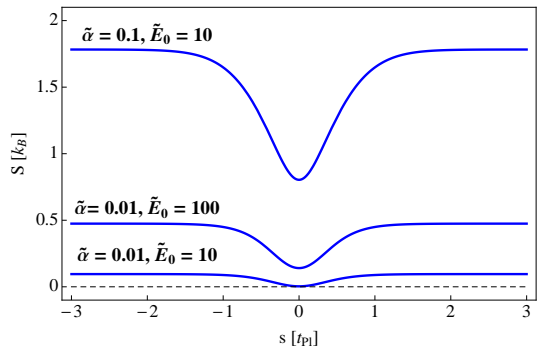}
\caption{Entropy in the bouncing cosmology.} \label{Fig2}
\end{figure}

In Fig. \ref{Fig2}, we present an evolution of the entropy  (\ref{DefEntropy})
for our system.
The entropy reaches the minimum value at the bounce while saturated
for $s\rightarrow \pm \infty$. This behavior is qualitatively similar to what was
predicted by the von Neumann entropy with environmental degrees of
freedom taken into account \cite{Zeh,Kiefer:2005vd}.

Making use of the above reasoning, we can give an interpretation
to the fact that for the Gaussian state the total uncertainty
$\Delta \hat{Q} \Delta \hat{P}\sqrt{1-\rho^2}$ has a minimum at the
bounce: {\it quantum entropy} grows with the increase of volume.
Furthermore, if the second law of thermodynamics applies to the
entropy defined by equation (\ref{DefEntropy}), one could expect
that entropic arrows of time are directed out of the point of
bounce. In such a case, only the expansion of the Universe can be
perceived by the internal observer. Such an interpretation differs
from the standard understanding of the phase of bounce as
successive contraction and expansion. This standard point of view
is additionally supported by the presence of internal quantum time
played by covariance, as shown in the previous section.

\section{Relative fluctuations}

\begin{figure}[ht!]
\centering
\includegraphics[width=7cm,angle=0]{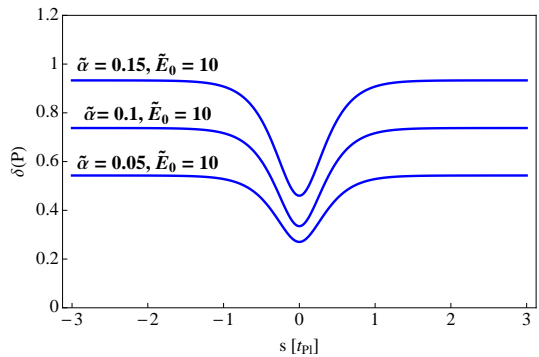}
\caption{Relative fluctuations of $\hat{P}$.} \label{Fig3}
\end{figure}

The relative {\it fluctuation} of an observable $\mathcal{O}$,
defined as $\delta(\mathcal{O}) :=\Delta
\hat{\mathcal{O}}/{\langle \hat{\mathcal{O}}} \rangle$ (where
${\langle \hat{\mathcal{O}}} \rangle$ is the expectation value of
$\mathcal{O}$), is a sort of a {\it gauge} for measuring the
cosmic amnesia.  The relative fluctuations $\delta(P)$ are
\emph{symmetric} with respect to the bounce, as shown in Fig.
\ref{Fig3}.
They  saturate on $\delta(P)|_{\text{max}}$ while $s\rightarrow
\pm \infty$ and reach the minimum at the bounce. Therefore, if the
semiclassicality condition $\delta(P)\ll1 $ is imposed in the
expanding phase, it constrains the rest of the evolution.

\begin{figure}[ht!]
\centering
\includegraphics[width=7cm,angle=0]{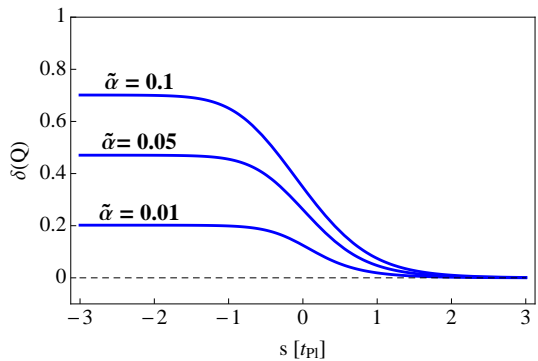}
\caption{Relative fluctuations of $\hat{Q}$.} \label{Fig4}
\end{figure}

In the case of the $\hat{Q}$ observable, the relative fluctuations
$\delta(Q)$, presented in Fig. \ref{Fig4},
are {\it asymmetric} with respect to the bounce and only become
symmetric when $ \tilde{\alpha} \rightarrow 0$. For $s \rightarrow
-\infty$, the $\delta(Q)$ saturates at $\delta(Q)|_{\text{max}}  =
\sqrt{e^{4\tilde{\alpha}}-1} $, while it tends to zero for $s
\rightarrow +\infty$. Since the directions of the evolution parameter
$s$ and cosmological time $t$ are opposite, the relative
fluctuations $\delta(Q)$ monotonically {\it grow} in the
cosmological time.

From the point of view of the possible {\it observability}
(detection) of the amnesia, the observable $\hat{Q}$ is a favorite
because in the classical limit, $Q = \gamma H$, where $H$ is the
Hubble parameter and $\gamma$ is Barbero-Immirzi parameter.
Therefore, relative fluctuations of $\hat{Q}$  {\it can be}
constrained observationally. In contrast,   it is hard to put any
constraint on $\delta(P)$, because  $P$  is linked to the physical
volume of space $v = 4\pi G \gamma P$, which is not measurable.

The relative fluctuations $\delta(Q)$ at the given time cannot be
greater than the relative uncertainty of measurement. In
particular, the present value of the Hubble factor is
$H_0=70.2\pm1.4 \ \text{km} \ \text{s}^{-1}\ \text{Mpc}^{-1}$
\cite{Komatsu:2010fb}, thus we have the constraint $\delta(Q) <
\frac{\sigma(H_0)}{H_0} \approx 0.02$. Another constraint can be
derived for the phase of inflation, based on observations of the
cosmic microwave background (CMB) radiation. If the inflation was
driven by a massive inflaton field, the Hubble factor $H_*
=\frac{1}{2} m_{\text{Pl}} \sqrt{ \mathcal{A}_s \pi (1-n_s)}$,
where $\mathcal{A}_s$ is the amplitude of scalar perturbations and
$n_s$ is the corresponding spectral index. From the seven years
of observations of the WMAP satellite combined with other
cosmological measurements $\mathcal{A}_s = (2.430\pm 0.091) \cdot
10^{-9}$ and $n_s =0.968\pm0.012$  \cite{Komatsu:2010fb}. Based on
this we find the constraint $\delta(Q) < \frac{\sigma(H_*)}{H_*}
\approx 0.19$. The future measurements of the B-type polarization
of the CMB will allow to determine $H_*$ with higher precision and
therefore improve the above constraint.

The model we consider applies to the vicinity of Planck's epoch,
however if assuming that the quantum fluctuations are not
decreasing thereafter, the derived observational bound can be used
to constraint  $\delta(Q)|_{\text{max}}$. From the first
constraint $\delta(Q)|_{\text{max}} < 0.02$, which translates into
$\tilde{\alpha} <10^{-4}$ and from the second one
$\delta(Q)|_{\text{max}} < 0.19$, leading to $\tilde{\alpha} < 9
\cdot 10^{-3}$. Both constraints suggest that the semiclassicality
condition was indeed fulfilled. Therefore, because constraint
$\delta(Q)|_{\text{max}}<1$ implies  $\delta(P)|_{\text{max}}<1$,
one can conclude that there is no cosmic amnesia within the considered
model.

\section{Summary and conclusions}

In this paper, we have studied quantum dynamics, using the Gaussian
state, of the FRW cosmological model with a free scalar field in
the framework of the reduced phase space loop quantum cosmology.

We have analyzed evolution of the first and second order moments
of the canonical variables. By investigating quantum uncertainties,
we have shown that the phase of the bounce is the least quantum
part of the evolution. We have also shown that covariance is a
monotonic function of the time parameter.  Therefore, it may serve
as an intrinsic quantum parameter of time.

We have introduced the notion of the entropy of squeezing and analyzed
its behavior for the the quantum state under considerations. We
have shown that the resulting scale factor dependence of entropy
is in qualitative agreement with the results based on the von Neumenn
entropy for the mixed states.

We have shown that the $\hat{Q}$ observable, not the $\hat{P}$ observable,
should be used as a tool for studying (within the FRW model) the reality of
the cosmic forgetfulness. It is so because only relative fluctuations of
$\hat{Q}$ can be constrained observationally, contrary  to the
fluctuations of $\hat{P}$.   The available cosmological data allow one 
to constrain the relative fluctuations  $\delta(Q)$. Using these
data, we have shown that the semiclassicality is preserved across
the bounce.

Since the dependance of $\Delta \hat{Q} \Delta \hat{P}\sqrt{1-\rho^2}$
on time is symmetric with respect to the bounce, our quantum universe is
equally quantum before and after the bounce, in the context of uncertainty
principle. In this sense, the Universe remembers its  quantumness
across the bounce.

In papers \cite{BojowaldNature,Bojowald:2007gc}, the authors consider a
solvable toy model (motivated by the LQC) to argue that a quantum
state before the bounce may become semiclassical after the bounce.
The authors of \cite{Corichi:2007am,Corichi:2011rt} criticize
these results claiming that the cosmic amnesia is only an artifact
of poor analyzes of a simple toy model. In \cite{Corichi:2007am,Corichi:2011rt},
the authors examine the same cosmological model as we do, but within the
sLQC method (simplified LQC). However, they mainly examine the $\hat{P}$
observable. Since $\delta(P)$ is symmetric with respect to the
bounce, they obviously cannot see any indications of the cosmic
amnesia. Also the sophisticated calculations presented in
\cite{Kaminski:2010yz} concern mainly $\delta(P)$.

Further analysis can be done by calculating an evolution of
$\delta(Q)$ for the FRW model with a scalar field {\it potential}.
This would enable obtaining more accurate  constraints from the
CMB observations.

We are conscious that our results may depend on the choice of
the initial state. Therefore, we have been working on the extension 
of our investigation considering a variety of semiclassical states \cite{JKW}.
For such states, the mean values follow the classical trajectories 
as in the case of the Gaussian state. However, evolution of quantum 
fluctuations may differ from the case described here. In particular, 
one can construct a squeezed vacuum state for which dispersions
of elementary variables remain constant during evolution, but the 
covariance is varying in time. 
  
Our classical Hamiltonian is unique for a given choice of an evolution 
parameter (time). But, it is commonly known that quantization of an 
observable may suffer from ambiguities. In our next paper \cite{JKW}, 
we consider different factor ordering of elementary variables defining 
the Hamiltonian to test the sensitivity of our results to this procedure.

The issues raised above require performing extensive calculations, 
which are beyond the scope of the present paper.

An extension of our results to the Bianchi type universes
\cite{Dzierzak:2009dj,Malkiewicz:2010py} is another direction in
our cosmology program. Some evidence indicating possible
relevance of  anisotropic effects is connected with  an observed CMB
anomaly called the ``axis of evil" \cite{Land:2005ad}. This is another
motivation, apart from the BKL results, for  applying  the {\it
homogeneous} models to describe the very early Universe.

\acknowledgments

J.M. has been supported by the Foundation of Polish Science.


\begin{thebibliography}{99}
\bibitem{conv} We choose $\;\hslash = 1 = c,\;$ so $\;t_{\text{Pl}} =
\sqrt{G}=1/m_{\text{Pl}}\;$, except where otherwise noted.
\end{thebibliography}
\end{document}